# Introduction of A Hybrid Monitor for Cyber-Physical Systems

J. Ceasar Aguma
Department of Computer Science
University of California
Irvine, CA USA
jaguma@uci.edu

Bruce McMillin
Department Of Computer Science
Missouri University Of Science
and Technology
Rolla, MO USA
ff@mst.edu

Amelia Regan
Department of Computer Science
University of California
Irvine, CA USA
aregan@ics.uci.edu

## ABSTRACT

Computing systems and mobile technologies have changed dramatically since the introduction of firewall technology in 1988. The internet has grown from a simple network of networks to a cyber and physical entity that encompasses the entire planet. Cyber-physical systems(CPS) now control most of the day to day operations of human civilization from autonomous cars to nuclear energy plants. While phenomenal, this growth has created new security threats. These are threats that cannot be blocked by a firewall for they are not only cyber but cyber-physical. In light of these cyber-physical threats, this paper proposes a security measure that promises to enhance the security of cyber-physical systems. Using theoretical cyber, physical, and cyber-physical attack scenarios, this paper highlights the need for additional monitoring of cyber-physical systems as an extra security measure. Additionally, we illustrate the efficiency of the proposed monitor using a Shannon entropy proof, and a multiple security domain nondeducibility (MSDND) proof.

## 1 Introduction

In May of 2017, a ransomware attack held most of the developed world hostage, crippling healthcare systems, manufacturing systems, and multiple critical infrastructures across the globe. The British National Health Services was forced to limit health care to only emergency cases [7]. If not for a timely kill switch, the attack could have brought forth catastrophic damage to nuclear plants, air transportation systems, and many other infrastructures. The Wannacry [7] ransomware attack is a recent example of a now critical threat. A great many cyber and physical attacks keep cropping up all over the world, most notably; the Iran stuxnet attack [4], which, according to Iran's civil defense agency, was still a threat in October of 2018 [18], the byzantine replay attack [19], and the Ukraine power grid attack that left more than 230,000 people without electricity [20]. [23] provides an extensive list of typical security threats that are facing smart-city CPS and detailed countermeasures available to defend against these. Because cyber-physical systems (CPS), are physical entities with cyber functionality, traditional cybersecurity measures are simply not sufficient to mitigate the threat posed by this new wave of cyber-physical attacks [1]. While traditional cyber attacks were easily deducible and susceptible to prevention by means of a firewall or antivirus software, it has been shown that recent attacks like the Iran Stuxnet attack could go undetected for long periods of time [5]. A search for a solution to these threats should, therefore, focus on making the occurrence of such attacks almost impossible, and if the attacks remain possible, then they should at least be swiftly deducible.

The protection of cyber-physical systems cannot depend on the effectiveness of a single detection mechanism [5]. However, the majority of the proposed Cyber-physical security measures have centered around the notion of a single monitoring unit. The Shadow Security Unit(SSU) [16] proposed by Cruz et al is a viable idea, but considering that the SSU is a single unit that employs only cybersecurity measures, a cyber attack that targets the central monitoring unit itself, if not detected early, could be fatal to the rest of the CPS. Scaglione, Peisert, and McParland acknowledge the need for both a centralized and distributed monitor but the proposed monitor is only an algorithm [14]. While it's a great algorithm, it's still a cyber measure which will inevitably be vulnerable to some cyber attack. The same could be said about the Intrusion Detection Systems(IDSs) [15], that is, IDSs are also a single cyber measure. For an extensive look at the many cyber attacks, industry CPS models, and common cyber measure, we direct the reader to [19].

Other than purely cyber measures, some scholars have proposed the use of physics based measures to detect attacks, but these very rarely provide ways to mitigate the attacks or prevent them in the first place [21]. The primary idea of those methods is that physical properties of the system models can be used to detect attacks. Their paper presents a detailed survey of recent physics-based attack detection schemes in CPS models. Our research proposes the addition of a hybrid monitor spread over virtual nodes with randomized features. This addition to a CPS would provide a much-needed auxiliary layer of security and also enhance attack deductibility.

This paper is arranged as follows; Section 2 introduces the tools used in testing the viability of the proposed hybrid monitor as a security measure for CPS. Section 3 gives a comprehensive look at the hybrid monitor, listing its features and the reasoning behind each feature. Section 4 explains the methodology used to demonstrate the efficiency and effectiveness of the hybrid monitor. Section 5 presents the test scenarios and proofs. Section 6 wraps up with a short conclusion.

## 2 Background

This research employs the Future Renewable Electric Energy Delivery and Management (FREEDM) [2] System as a model CPS. Shannon entropy [8] is used as a tool to test the effectiveness of the monitor as a security measure. The multiple security domain nondeducibility(MSDND) [1] is used as a tool to test the effectiveness of the monitor in detecting attacks.

The FREEDM system center is a engineering research center funded by the National Science Foundation and spanning number of universities including but not limited to North Carolina State University, Missouri University of Science and Technology and Florida State University. The research center developed an energy management and distribution smart grid system located at the North Carolina State University that is also referred to as FREEDM [2]. The heart of this energy system is the Distributed Grid Intelligence (DGI) [2], an intelligent algorithm that implements energy management and distribution using modular adapters to interact with devices in a smart grid over different interfaces. In this paper, the FREEDM system is employed as a test subject for the implementation of the hybrid monitor proposed by this research.

The MSDND model is a security model that tests the integrity and confidentiality of a cyber-physical architecture. The MSDND model uses logic proofs to test information flow security; that is, how information moves among user groups within the security domains (SDs) that make up the system [1]. Howser and Mcmillin show that maintaining information flow security in CPS is challenging because the flow is irrevocably linked among the CPS' cyber and physical units [1]. Therefore, the MSDND model is employed to account for both cyber and physical information flow paths. The model defines two system properties namely: MSDND secure and notMSDND secure [1]. An MSDND secure system implies that for information flowing from entity A to entity B, entity B cannot deduce whether the information is valid or erroneous. This also means while an MSDND secure system is desirable if the goal is to maintain confidentiality, it can be an indicator of the possibility of an attack going undetected. A notMSDND secure system implies the alternative, that is, entity B can evaluate the correctness of information obtained from entity A. Therefore, in the event of an attack, Howser, and McMillin shows that a notMSDND secure system could easily detect the occurrence of an attack [1]. Thudimila and McMillin demonstrate the superiority of MSDND over traditional electronic and cryptographic solutions [3] when applied to detection of attacks in Automatic Dependent Surveillance-Broadcast(ADS-B) air traffic surveillance system [3]. For this research, MSDND is used to analyze whether a CPS with a hybrid monitor in place would deduce the occurrence of an attack.

Phan et al introduce the idea of using information flow-metrics like Shannon entropy to measure information leakage in CPS programs [9]. Li uses Shannon Entropy to break down the physical dynamics of CPS and goes on to show the negative entropy that communication adds to the general entropy of a CPS [10]. In this text, Shannon Entropy is used to illustrate the decrease in the possibility of an attack in a CPS after the introduction of a hybrid monitor. This demonstrates the value of adding a hybrid monitor to CPS architectures as an extra security measure.

Shannon entropy is an information theory concept derived from the general idea of information entropy that was developed and introduced by Claude Shannon [8]. Entropy is basically a measure of uncertainty in a communication system where a low entropy value implies minimal uncertainty and a high entropy implies the contrary. Shannon entropy defines entropy (***H***) as:

$$H[X] = E[I(X)]$$

Where ***I*** is the information content of the discrete random variable ***X***. Therefore we can further define *E[I(X)]* in terms of the probability mass function of X: i.e,

$$E[I(X)] = E[-log(P(X))] = -\sum_i P(x_i)log[P(x_i)]$$

The generally definition of entropy H then comes to:

$$H[X] = -\sum_i P(x_i)log[P(x_i)]$$

## 3 The Hybrid Monitor

To better protect cyber-physical systems, this research proposes the addition of a monitor. Many other researchers have explored the use of hybrid monitoring to ensure safety or/and security in CPSs. Li et al. propose extended hybrid automata modeling for vehicular CPSs as a safety and control measure [11]. Mao and Chen also introduce a runtime hybrid automaton monitoring framework for the Cooperative Adaptive Cruise Control Systems(CACC) [12]. This research borrows the idea of hybrid monitoring but with randomization as an additional feature. The addition of randomization in information flow paths' generation increases the system's entropy and in turn, reduces the chances of a successful attack in a generic CPS. That is because a higher number of information flow paths increases the number of points the attacker has to corrupt to remain undetected. Below is a detailed break down of this hybrid monitor's features;

- The monitor is hybrid, that is, both virtual and physical, central and decentralized. The monitor would have both a virtual component and physical component. The virtual components would be implemented as a hidden algorithm in every Supervisory control and data acquisition unit (SCADA) in the CPS. The physical component of the monitor would be a physical unit independent of the entire CPS and running a monitoring algorithm whose function is to oversee the operations of the monitor's virtual components. The physical component is, therefore, a central unit and the virtual components are the decentralized units.

- The monitor should be intelligent enough to generate physical invariants for every information flow path in the CPS. An invariant is simply a logical assertion that should always be true throughout an execution cycle. Therefore physical invariants are logical properties of a CPS that cannot be transformed by cyber entities and should always be held true. Having physical invariants makes the CPS vastly more secure because they are secure from being corrupted by cyber attacks. With that in mind, the hybrid monitor uses generated physical invariants as a validator of the information received from other system modules. The automated generation of physical invariant using machine learning, deep learning or linear regression is also a viable research area. The automation of invariant generation is explored further by Cruz et al [16].

- The virtual components of the monitors would continuously generate a randomly increasing number of paths for the flow of information between any two CPS entities. The physical monitor should also generate a randomly increasing number of virtual paths as a compliment to a physical path for the flow of information between any two virtual components of the monitor.

- All paths generated by the monitor should be independent of each other. This ensures that all the randomly generated paths cannot be collectively corrupted by an attacker.

- To reduce the information flow overhead, information sent through the monitors should be sent through a randomly chosen path among the generated paths and then white noise should be transmitted on the rest of the paths.

- The monitor should have a routing algorithm that can be employed if the monitor detects a failure or corruption at any of the CPS' entities.

- Communication between the virtual and physical monitor should be done on an entirely different network than that used by the rest of the cyber-physical system.

Note that this hybrid monitor is only a theoretical idea but the exact physical realization should at the very least aim to implement the above mentioned features.

## 4 Methodology

This research uses two methods to highlight the significance of introducing a hybrid monitor to a CPS.

### 4.1 Method 1

The research employs attack scenarios to examine the security of a CPS with and without the hybrid monitor. There are three attack scenarios, that is; A purely cyber attack like a ransomware on a CPS, a completely physical attack like the attacker inflicting physical damage to the CPS by, for example, cutting wires and a cyber-physical attack like the Iran Stuxnet attack expounded upon in Kushner's [4] and Karnouskos' [5] work.

### 4.2 Method 2

In the second method, the research uses two proof models i.e Shannon entropy and MSDND to show that the addition of a hybrid monitor makes a CPS less susceptible to an undetected attack and much more effective at deducing attacks when they do occur.

## 5 Research Results

This section details the three attack scenarios, their respective results, the Shannon entropy proof, and MSDND proof.

### 5.1 Cyber Attack scenario

As mentioned in the background, the FREEDM system is controlled by a distributed algorithm called the DGI. The DGI is set up to run on multiple nodes spread out over a network. It provides an interface for energy management applications to communicate with physical power devices.

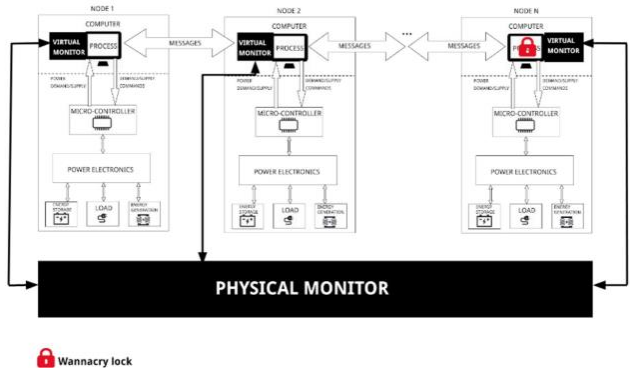

**Figure 1: The DGI under a WannaCry attack**

Let us assume that a DGI node is being held hostage by the wannaCry ransomware (figure 1). This kind of attack is rather easy to detect because ransomware attacks normally make the user aware that the attack is in progress. Therefore, for this scenario, attack deductibility is not important. But because the attacker is holding a node hostage, all information flowing through this node could be infected by the attacker. This could give the attacker further access to other nodes since all nodes of the DGI share state information. At this point, it's clear that the entire DGI could be held hostage. Since the DGI manages the entire FREEDM smart grid system, the entire CPS would be either rendered useless or could be left vulnerable to more damaging attacks.

Now let's consider a scenario where a hybrid monitor was in place with virtual units running alongside every DGI node and a physical unit to oversee the virtual units. Because all traffic that goes through a node is verified by the monitor and subjected to physical invariants generated by the monitor, it would be easy for the monitor to flag the presence of the ransomware. Since the monitor has information flow routing capabilities, all state information from other nodes would be safely rerouted through other nodes. While the infected node would not be saved, the rest of the DGI would continue to function without threat.

### 5.2 Physical Attack scenario

For this scenario, we assume that an attacker has inflicted physical damage to the CPS without using cyber means. The damage could be as simple as cutting an Ethernet cord or breaking a sensor. The detection and solution for such an attack are also rather simple. However, if the CPS is a critical infrastructure like a nuclear reactor that needs to continuously keep some functions fully operational then even this simple attack could prove fatal. With a monitor in place, any failure in the CPS would quickly be detected. The monitor, through information flow rerouting, would go even further to keep critical functions running while the damage gets fixed.

### 5.3 Cyber-physical Attack scenario

The third and last scenario assumes that a microcontroller in FREEDM system is infected by Stuxnet (figure 2). Erroneous Information from this microcontroller could cause catastrophic damage to the smart grid. In this case, deducing the presence of the Stuxnet and reducing the damage to the smart grid are both necessary.

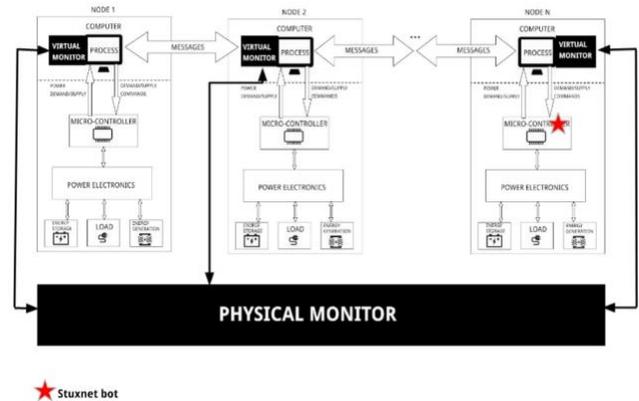

**Figure 2: The DGI under a Stuxnet attack**

From the Iran attack, it's clear that the Stuxnet could go unnoticed for a long time if no extra security measure is put in place. Although, if the FREEDM system had a hybrid monitor, the Stuxnet would be detected because all information from the microcontroller would have to be verified by the monitor. Since the monitor has physical invariants to prove the correctness of information from this microcontroller, any discrepancies in the information generated by the Stuxnet would be caught. On detection, information flow would then be routed through other nodes and further infection would be avoided. The Stuxnet would have to infect all random paths used by the hybrid monitor to avoid detection. The Shannon Entropy proof below shows that there is a very small possibility of the Stuxnet or attacker infecting all of the hybrid monitor's random paths.

### 5.4 MSDND proof

For this proof, let's look at the cyber-physical Stuxnet attack shown above. More specifically, the information path between the infected microcontroller and the DGI node process running on the computer without the monitor.

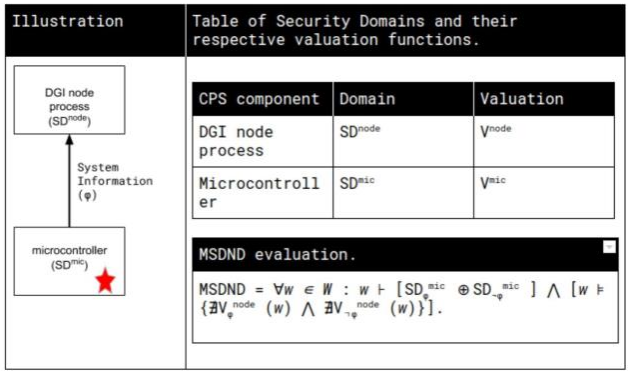

**Figure 3: MSDND evaluation for the DGI without a monitor**

Let us define the two domains as SDnode for the DGI node and SDmic for the microcontroller with valuation functions Vnode and Vmic respectively. Then consider a scenario where arbitrary information ($\varphi$) is sent from the infected microcontroller to the DGI node process. If the DGI node process and microcontroller are at the same level of security, then the DGI node process will trust that information from the infected microcontroller to be valid. Since the information can be either true or false, the first condition; i.e,.. (SD$\varphi$mic, SD¬$\varphi$mic ) for MSDND is met [1]. This is derived from the fact that if $\varphi$ is true then SD$\varphi$mic is true or if $\varphi$ is false then SD¬$\varphi$mic is true hence the xor statement is always true.

The second condition is also satisfied from the assumption that the two domains are at the same security level [1]. Therefore, the DGI node process believes and trusts the infected microcontroller. This means the DGI node process has no valuation function to prove the validity of $\varphi$ [1]. The absence of this valuation function (V$\varphi$node) leaves the system in an MSDND secure system [1]. This is the MSDND secure evaluation shown in figure 4.

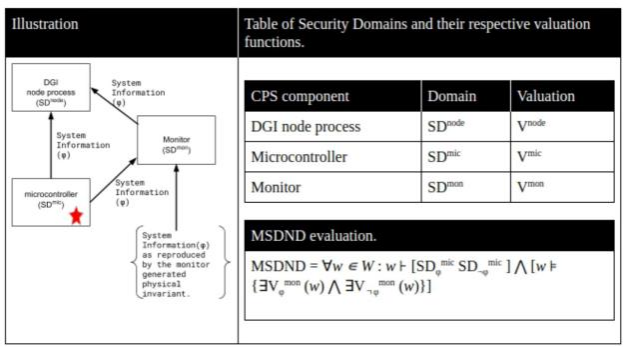

**Figure 4: MSDND evaluation for the DGI with a monitor**

The implication of this MSDND evaluation is that if the infected microcontroller sent false information to the DGI node process, there would be no way of evaluating that the information is false. Therefore the Stuxnet would go undetected. Knowin this, let us take a look at a scenario with the monitor in place.

The difference in this scenario is the presence of a monitor that is equipped with physical invariants. Using a physical invariant, the monitor can evaluate the validity of $\varphi$. With this, the monitor can also determine the state of the microcontroller with respect to the validity of $\varphi$; i.e,.. There exist a valuation V$\varphi$mon leaving the state SD$\varphi$mic deducible [1]. Hence the notMSDND secure evaluation shown in figure 4.

The proof shows that in the event of a cyber-physical attack like the Stuxnet attack, the presence of a monitor would render the attack deducible. For the attack to go undetected with a monitor in place, the attacker would have to infect every single monitor node, both virtual and physical. The next Shannon entropy proof will show that the possibility of compromising all the monitor nodes without being detected is rather minimal.

### 5.5 Shannon Entropy proof

The proof considers two scenarios where the attacker is attempting to infect the information flow between the DGI node process and the microcontroller.

First, let us take a look at the entropy of the setup without the monitor. There are two possible information flow events $x\_1$ and $x\_2$ that the attack could target. With a sample space = 2, the probability of the attacker successfully infecting information flow between the DGI node process and microcontroller is ½. The entropy evaluation for this scenario is shown below.

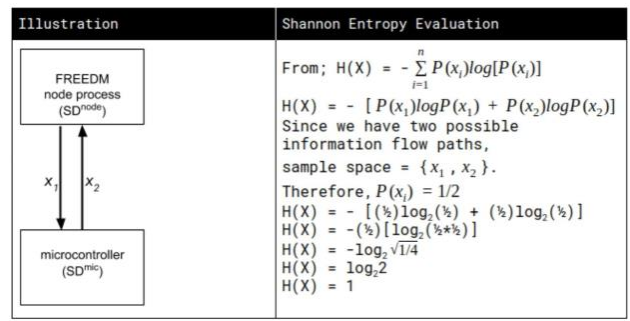

**Figure 5: Shannon Entropy evaluation for the DGI without a monitor**

With a monitor in place, the sample space grows to (2n + 2), making the probability of successfully corrupting one path come to 1/[2(n + 1)]. Here is the entropy evaluation;

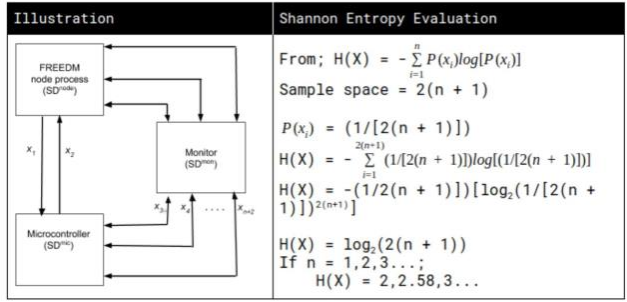

**Figure 6: Shannon Entropy evaluation for the DGI with a monitor**

The proof shows us that the entropy increases with the increase in the size of n paths. From the attacker's point of view, the uncertainty increases with increasing size of n paths. Therefore as the size of n increases, it becomes much harder for the attacker to launch a successful attack on the CPS. By adding the hybrid monitor, the system is not fully secure from an attack but the possibility of a successful attack is vastly smaller.

## 6   Conclusion

After 2017's Ransomware attack [17], the world can not ignore the threat posed by the possibility of using attacks on cyber-physical systems as a tool for terrorism and cyber warfare. The increased occurrence of cyber-physical systems attacks is surely an indicator that traditional cybersecurity measures are insufficient at prevention and detection of these attacks. The world needs to start considering alternative or improved security measures. The combination of an intelligent, randomized physical monitor with existing virtual cyber measures to create a hybrid monitor is a good place to start. While, the hybrid monitor is not a foolproof solution to cyber-physical attacks, it could well be the best solution yet. Future research in this area should focus on physical implementation of the hybrid monitor and prevention of attacks targeting the hybrid monitor itself. This hybrid monitor could be the great leap towards fully securing an important and nonexpendable entity of smart living that is cyber-physical systems.

## ACKNOWLEDGMENTS

This research was sponsored by the United States National Science Foundation (NFS). The following colleagues made notable contribution to the research over the span of the project; Dr. Patrick Taylor (Associate Professor, Missouri S&T's Department of Computer Science), Manish Jaisinghani (Graduate Student, Missouri S&T's Department of Computer Science), Anusha Thudmilla (Graduate Student, Missouri S&T's Department of Computer Science),Joshua Hermann (Graduate Student, Missouri S&T's Department of Computer Science).